\documentstyle[aps]{revtex}
\begin{document}
\centerline{\LARGE\bf Interpreting the wave function and its
evolution} \centerline{\LARGE\bf in terms of new motion of
particle} \vskip 1cm \centerline{Gao shan} \centerline{Institute
of Quantum mechanics} \centerline{11-10, NO.10 Building, NO.21,
YueTan XiJie DongLi, XiCheng District} \centerline{Beijing 100045,
P.R.China} \centerline{E-mail: gaoshan.iqm@263.net}

\vskip 1cm
\begin{abstract}
Aiming at providing an objective motion picture for the
microscopic object described by the wave function, new analysis
about motion is presented by use of the point set theory in
mathematics, through which we show that a new kind of motion named
quantum discontinuous motion is the general motion mode of the
particle, while classical continuous motion is just one kind of
extremely peculiar motion, and the wave function in quantum
mechanics proves to be the very mathematical complex describing
the particle undergoing the quantum discontinuous motion.
Furthermore, Schr\"{o}dinger equation of the wave function is
shown to be the simplest
nonrelativistic evolution equation for the particle undergoing the
new motion, and the consistent axiom system of quantum mechanics
is also deduced out. At last, we demonstrate that present quantum
measurement theories just confirm the existence of the new motion
of the microscopic particle described by the wave function, and
the weird displays of the wave function in microscopic world are
also physically explained in terms of the new motion.
\end{abstract}

\vskip 1cm

\section{Introduction}
Owing to the springing up of the new concepts about the wave
function such as quantum cosmology and protective
measurement\cite{Hawking}\cite{Aha1}\cite{Aha2}, the demand for
the last objective interpretation of the wave function is
skyrocketing in recent years. On the other hand, the analyses
about motion have never ceased since the old Greece times, but
from Zeno Paradox to Einstein's relativity\cite{Einstein}, only
classical continuous motion is discussed. In fact, there exist
some other motion modes in nature, and the mysterious quantum
motion in microscopic world may be one of these snubbed motion
modes. In this paper we present one kind of new motion, which is
called quantum discontinuous motion, and after a deep and
reasonable analysis we find that the last reality underlying the
wave function is just the new motion undergone by the microscopic
particle.

From a historical point of view, interpreting the wave function in
terms of new motion of particle inherits the essence of thought
accumulated since the foundation of quantum mechanics, the first
essential concept is the ontology concept widely adopted in
interpreting quantum mechanics, it aims at finding the last
reality behind the mysterious wave function, this concept can be
traced back to Einstein, Schr\"{o}dinger, and the followers
Bohm\cite{Bohm}, Bell\cite{Bell} et al, especially in recent
years, the protective measurement\cite{Aha1}\cite{Aha2} presented
by Aharonov et al has not only consolidated this concept from the
inside of quantum mechanics, but also provided the experiments to
confirm it, thus this advance cries for an objective elucidation
about the reality underlying the wave function more strongly than
ever, and it may be the time to disclose the whole quantum mystery
now.

The second is the particle concept hold by Born's
generally-accepted probability interpretation\cite{Born} and
nearly most following interpretations of quantum mechanics,
according to this particle concept the last reality described by
the wave function in quantum mechanics is essentially a particle,
which is in only one position in space at any instant. The most
predominant character of this concept is that it can not only
describe the particle picture in nonrelativistic quantum
mechanics, but also be extended to describe the field picture in
relativistic quantum field theory when uniting new motion with
special relativity; on the other hand, it provides a precondition
to further study the boring collapse process as one kind of
objective process.

The third is the renunciation of classical continuous motion, for
example, it is widely demonstrated and accepted by ontologists
that the microscopic particle can not pass through only one slit
without influenced by another slit in two-slit experiment, and
sometimes they even say that the microscopic particle "passes
through both slits" or "in two positions at the same
time"\cite{Aha1}\cite{Aha2}, although this kind of description is
too vague to form a strict definition or elucidation about the
objective quantum motion undergone by the microscopic particle,
which may be extremely different from classical continuous motion,
needless to say, the renunciation of classical continuous motion
sheds light on the road to interpret the wave function in terms of
new motion other than classical continuous motion, and the vague
description "in two positions at the same time" indeed grasps
something real and moves a peg along this road.

Certainly, there exist other deeper logical reasons leading to
interpreting the wave function in terms of new motion of particle,
some of them are as follows:

(1). The linear superposition principle of wave function in
quantum mechanics strongly implies that the real object described
by the wave function, if any, is not a field, but a particle,
since as to the field its different branches will generally
interact, and the linear superposition principle of the wave
function describing the field will be broken; while as to the
particle, it is in only one position in space at any instant, and
in nonrelativistic quantum mechanics the transfer of interaction
is instantaneous, thus for the wave function describing the
quantum discontinuous motion of particle, its different branches
will not interact.

(2). Since the founding of science people have been studying the
classical continuous motion of particle, and taking it for granted
that it is the only type of motion in Nature, but from the
mathematical analyses it is shown that there exists one kind of
complementary motion mode of classical continuous motion, which
can be called quantum discontinuous motion, they form the complete
set of motion modes of particle, and the former seems more natural
than the latter. In physics, as we will demonstrate, the
macroscopic world is governed by the classical continuous motion,
while the microscopic world is governed by the quantum
discontinuous motion.

(3). In order to unify particle and field, we have been searching
for the lost connection between them, now the new motion---quantum
discontinuous motion is just the lost golden bridge. On the one
hand, the object undergoing the new motion is a particle, this is
the particle-like aspect of the new motion, owing to this
particle-like property, the new motion of particle provides a
essential basis to account for the objective collapse process of
the wave function during quantum measurement, since the definite
measurement result such as one spot on the photoplate, is produced
only in one local region, not in many different local regions in
space; On the other hand, the particle undergoing the new motion
moves throughout the whole space with a certain position measure
density $\rho(x,t)$ during infinitesimal time interval, this is
the field-like aspect of the new motion, owing to this field-like
property, the new motion of particle can describe the objective
evolution process of the wave function during normal evolution,
which determines the interference pattern of the particle, and
will provide the objective origin of probability in quantum
mechanics.

(4). When the new motion marries with special relativity, we will
find that it can be extended to describe the quantum field
picture, and quantum mechanics will naturally be taken over by
quantum field theory, concretely speaking, when in relativistic
quantum field the transfer velocity of interaction is finite, thus
for the wave function describing the quantum discontinuous motion
of particle, its different branches will interact through the
transfer particle of the interaction, for example, in quantum
electrodynamics the different branches of the electron wave
function will interact through the photon, which is characterized
by the interaction term $\bar \psi \gamma^{\mu} A_{\mu} \psi$.


(5). The new motion of particle will provide a broader framework
for objectively studying the microscopic process, in which the
notorious collapse problem may be naturally solved, for example,
there may exist many kinds of concrete motion modes among the new
motion, and the new motion may display differently in the
nonrelativistic and relativistic domains, especially when
involving relativistic gravity the new motion of particle may
naturally provide the origin of randomness in the collapse
process, and further result in the objective collapse process.

(6). It is generally accepted that the main accepted objective
interpretations of quantum mechanics are Everett's relative state
interpretation\cite{Everett} and Bohm's hidden variables
interpretation\cite{Bohm}, but besides other unsatisfactory
characters they all interpret neither the wave function, nor
Schr\"{o}dinger equation of the wave function, as to the former,
the wave function is directly taken as a physical entity with no a
priori interpretation, while as to the latter, the wave function
is taken as a real field in the configuration space, which is
assumed to satisfy Schr\"{o}dinger equation in quantum mechanics
with no further interpretation, in fact, it does not interpret
this objective field in real space either. Now, in this paper we
will interpret both the wave function and its Schr\"{o}dinger
equation in terms of the new motion of particle.

The plan of this paper is as follows: In Sect. 2 we first give
three general presuppositions, which relate physical reality with
abstract mathematics, and are the basis of the following analysis
about motion. In Sect. 3 we give a strict mathematical analysis
about motion by use of point set theory, especially the
discontinuous motion described by regular dense point set is
analyzed in detail. In Sect. 4 a strict physical definition of the
new motion---quantum discontinuous motion is given, the wave
function in quantum mechanics is interpreted as a mathematical
complex describing the particle undergoing the quantum
discontinuous motion, and the consistent axiom system of quantum
mechanics is deduced out, especially Schr\"{o}dinger equation of
the wave function is shown to be the simplest nonrelativistic
evolution equations for the particle undergoing the quantum
discontinuous motion. Furthermore, in Sect. 5 we demonstrate that
the quantum measurement theories just confirm the existence of the
quantum discontinuous motion of the microscopic particle described
by the wave function. At last, in Sect. 6 the notorious characters
of wave function is consistently interpreted, and two concrete
examples are given to explain the weird displays of the wave
function in microscopic world in terms of the quantum
discontinuous motion of particle.

\section{Three general presuppositions}
First, we will give three general presuppositions about the
relation between physical reality and abstract mathematics, which
are basic conceptions and correspondence rules before we discuss
the physical motion of particle, we suppose they are the
commonness for all kinds of physical motion of particle.

(1). Time and space in which the particle moves are both
continuous point set.

(2). The moving particle is represented by one point in time and
space.

(3). The motion of particle is represented by the point set in
time and space.

For simplicity but lose no generality, in the following we will
mainly analyze one-dimension motion, namely the point set in
two-dimension time and space.

\section{Mathematical analysis about new motion}

\subsection{Point set and its law}

As we know, the point set theory has been deeply studied since the
beginning of this century, nowadays we can grasp it more easily.
According to this theory, we know that the general point set is
dense point set, whose basic property is the measure of the point
set, while the continuous point set is one kind of particular
dense point set, its basic property is the length of the point
set.

Naturally, as to the point set in two-dimension time and space,
the general situation is the dense point set in this two-dimension
space, while the continuous curve is one kind of extremely
particular dense point set, surely it is a wonder that so many
points bind together to form one continuous curve by order, in
fact, the probability for its formation is zero.

Now, we will generally analyze the law of the point set, as we
know, the law about the points in point set, which can be called
point law, is the most familiar law, and it is widely taken as the
only rational law, for example, as to the continuous curve in
two-dimension time and space there may exist a certain expressible
analytical formula for the points in this particular point set(
people cherish this kind of point laws owing to their infrequent
existence, but perhaps Nature detests and rejects them, since the
probability of creating them is zero), but as to the dense point
set in two-dimension time and space the point law does not exist,
since the dense point set is discontinuous everywhere, even if the
difference of time is very small, or infinitesimal, the difference
of space can be very large, then infinitesimal error in time will
result in finite error in space, thus even if it exists we can not
formulate it in nature, and owing to finite error in time
measurement, we can not confirm it either, let alone predict the
evolution of the point set using it, in one word, there does not
exist point law for dense point set in mathematics and physics.

Because of nonexistence of the point law for general dense point
set, people cherish only the particular dense point
set---continuous curve with point law, which corresponds to
classical continuous motion, and detest the general dense point
set without point law, let alone regard it as another kind of real
motion mode. But when we consider the confirmation of law, we will
find more truth about the law for point set, as we know, as to the
point law of continuous curve, we must confirm it by means of the
following process:$\Delta{t}\rightarrow{dt}\rightarrow0$, among
these processes the process $\Delta{t}\rightarrow{dt}$ is complete
for confirming the differential law for point set, and the process
${dt}\rightarrow0$ is only necessary for the confirmation of point
law, but evidently this process can not be achieved in reality, in
fact, only the process $\Delta{t}\rightarrow{dt}$ can possess real
physical meaning through testing the law more and more accurately,
thus there does not exist point law for both general dense point
set and continuous curve, and the privilege of continuous curve
and the corresponding classical continuous motion is also lost.

On the other hand, in physical there exist only the dynamical
quantities defined during infinitesimal time interval, this fact
can be seen from the familiar differential quantities such as dt
and dx, whereas the point quantities come only from mathematics,
people always mix up these two kinds of quantities, this is a huge
obstacle for the development of physics. Thus we can only discuss
the quantities defined during infinitesimal time interval, as well
as their differential laws if we study the point set corresponding
to real physical motion.

\subsection{Deep analysis about dense point set}
Now, we will further study the differential description of point
set in detail.

First, in order to find the differential description of the
peculiar dense point set---continuous curve, we may measure the
rise or fall size of the space $\Delta{x}$ corresponding to any
finite time interval $\Delta{t}$ near each instant $t_j$, then at
any instant $t_j$ we can get the approximate information about the
continuous curve through the quantities $\Delta{t}$ and
$\Delta{x}$ at that instant, and when the time interval
$\Delta{t}$ turns smaller, we will get more accurate information
about the curve. Theoretically we can get the complete information
through this infinite process, that is to say, in theory we can
build up the basic description quantities for the peculiar dense
point set---continuous curve, which are the differential
quantities dt and dx, then given the initial condition the
relation between dt and dx at all instants will completely
describe the continuous curve.

Then, we will deeply analyze the differential description of the
general dense point set. As to this kind of point set, we still
need to study the concrete situation of the point set
corresponding to finite time interval near every instant. Now,
when time is during the interval $\Delta{t}$ near instant $t_j$,
the points in space are no longer limited in the local space
interval $\Delta{x}$, they distribute throughout the whole space
instead, so we should study this new nonlocal point set, which is
also dense point set, for simplicity but lose no generality, we
consider finite space such as x$\in$[0,1], first, we may divide
the whole space in small equal interval $\Delta{x}$, then
calculate the measure of the local dense point set in the space
interval $\Delta{x}$ near each $x_i$, which can be written as
$M_{\Delta{x},\Delta{t}}$($x_i$,$t_j$), since the measure sum of
all local dense point sets in time interval $\Delta{t}$ just
equals to the length of the continuous time interval $\Delta{t}$,
we have:
\begin{equation}\sum_{i}M_{\Delta{x},\Delta{t}}$($x_i$,$t_j$)=
$\Delta{t}\end{equation}

On the other hand, since the measure of the local dense point set
in the space interval $\Delta{x}$ and time interval $\Delta{t}$
will also turn to be zero when the intervals $\Delta{x}$ and
$\Delta{t}$ turn to be zero, it is not an useful quantity, and we
have to create a new quantity on the basis of this measure.
Through further analysis, we find that a new quantity
$\rho_{\Delta{x},\Delta{t}}(x_i,t_j)=M_{\Delta{x},\Delta{t}}(x_i,t_j)/(\Delta{x}\cdot\Delta{t})$,
which can be called average measure density, will be an useful
one, it generally does not turn to be zero when $\Delta{x}$ and
$\Delta{t}$ turn to be zero, especially if the limit
$lim_{\Delta{x}\rightarrow0,\Delta{t}\rightarrow0}\rho_{\Delta{x},\Delta{t}}(x_i,t_j)$
exists, it will no longer relate to the observation sizes
$\Delta{x}$ and $\Delta{t}$, so it can accurately describe the
whole dense point set, as well as all local dense point sets near
every instant, now we let:
\begin{equation}
\rho(x,t)=lim_{\Delta{x}\rightarrow0,\Delta{t}\rightarrow0}\rho_{\Delta
{x},\Delta{t}}(x,t)\end{equation} then we can get:
\begin{equation}
\int_{\Omega}\rho(x,t)dx=1
\end{equation}
this is just the normalization formula, where $\rho$(x,t) is
called position measure density, $\Omega$ denotes the whole
integral space, we call this kind of dense point set regular dense
point set.

Now, we will analyze the new quantity $\rho$(x,t) in detail,
first, the position measure density $\rho$(x,t) is not a point
quantity, it is defined during infinitesimal interval, this fact
is very important, since it means that if the measure density
$\rho(x,t)$ exists, then it will be continuous relative to both t
and x, 
that is to say, contrary to the position function x(t), there does
not exist the discontinuous situation for the measure density
function $\rho$(x,t), furthermore, this fact also results in that
the continuous function $\rho$(x,t) is the last useful quantity
for describing the regular dense point set; Secondly, the
essential meaning of the position measure density $\rho$(x,t) lies
in that it represents the dense degree of the points in the point
set in two-dimension space and time, and the points are denser
where the position measure density $\rho$(x,t) is larger.


\subsection{The evolution of regular dense point set}
Now, we will further discuss the evolution law for regular dense
point set.

Just like the continuous position function $x(t)$, although the
continuous position measure density function $\rho$(x,t)
completely describes the regular dense point set, it is one kind
of static description about the point set, and it can not be used
for prediction itself, so in order to predict the evolution of the
regular dense point set we must create some kind of quantity
describing its change, enlightened by the theory of fluid
mechanics we can define the fluid density for the position measure
density $\rho$(x,t) as in the following:
\begin{equation}
\frac{\partial{\rho}(x,t)}{\partial{t}}+\frac{\partial{j}(x,t)}{\partial{x}}
=0
\end{equation}
we call this new quantity $j(x,t)$ position measure fluid density,
it is evident that this quantity just describes the change of the
measure density of the regular dense point set, thus the general
evolution equations of the regular dense point set can be written
as in the following:
\begin{equation}
\frac{\partial{\rho}(x,t)}{\partial{t}}+\frac{\partial{j}(x,t)}{\partial{x}}=0
\end{equation}
\begin{equation}
\frac{\partial{j(x,t)}}{\partial{t}}+f(\rho,j,\frac{\partial{\rho}}{\partial{x}},\frac{\partial{j}}{\partial{x}}
,...)=0
\end{equation}
where
$f(\rho,j,\frac{\partial{\rho}}{\partial{x}},\frac{\partial{j}}{\partial{x}}
,...)$ is a certain function containing $\rho(x,t)$,$j(x,t)$ and
their partial derivatives relative to x.

\section{Physical analysis about new motion}
In this part, we will return to the physical world and analyze the
physical motion.

\subsection{The definition of new motion}
Through the presuppositions presented in the beginning of this
paper, we will give a strict physical definition of new motion in
three-dimension space, the definition for other abstract spaces or
many-particle situation can be easily extended.

(1). The new motion of particle in space is described by regular
dense point set in four-dimension space and time.

(2). The new motion state of a particle in space is described by
the position measure density $\rho(x,y,z,t)$ and the position
measure fluid density $\mathbf{j}(x,y,z,t)$ of the corresponding
regular dense point set, in the simplest situation they form an
integrated abstract wave function
$\psi(x,y,z,t)=\rho^{1/2}\cdot{e^{iS(x,y,z,t) / \hbar}}$, where
$S(x,y,z,t)=m\int_{-\infty}^{x}j_{x}(x^{'},y,z,t) /
\rho(x^{'},y,z,t)dx^{'}$, $m$ is the mass of the particle and
$\hbar$ is a constant quantities.

(3). The evolution of new motion corresponds to the evolution of
regular dense point set, and one of its evolution equations
containing the wave function $\psi(x,y,z,t)$ is Schr\"{o}dinger
equation in quantum mechanics.

We call the new motion quantum discontinuous motion, compared with
classical continuous motion, the commonness of these two kinds of
motion is that the particle exists only in one position in space
at one instant, their difference lies in the behavior of the
particle during infinitesimal time interval [t,t+dt], for
classical continuous motion, the particle is limited in a certain
local space interval [v,v+dv], while for quantum discontinuous
motion, the particle moves throughout the whole space with a
certain position measure density $\rho(x,y,z,t)$.

On the other hand, in order to analyze the physical evolution of
the new motion, one point needs to be emphasized, according to the
above definition of new motion, the particle is in only one
position at any instant, but this is not a direct physical
statement, but a simple metaphysical statement about the
time-division existence of the
reality undergoing the new motion, which is logically deduced from
the experiment fact that the wave function satisfies the linear
superposition principle, or there does not exist self-interaction
for the wave function in quantum mechanics, strictly speaking, the
particle may be not in only one position at one instant( we can
not confirm it in physics either ), the only rational physical
requirement is that the instants set, at any instant of which the
particle is in only one position, forms a dense point set, whose
measure equals to the length of the time interval. In fact, the
following physical states of the new motion are all defined during
infinitesimal time interval in the meaning of measure, not at one
instant, for example, the momentum eigenstate
$\psi_{p}(x,t)=e^{ipx-iEt}$, especially even the position
eigenstate $\psi(x,t)=\delta(x-x_0)$ are still defined during
infinitesimal time interval.

\subsection{The evolution of new motion}
In the following, we will give the main clues for finding the
possible evolution equations of the new motion, at the same time,
the axiom system of quantum mechanics will be deduced out by use
of a logical analysis, and Schr\"{o}dinger equation will prove to
be just the simplest nonrelativistic evolution equations of the
new motion. For simplicity but lose no generality, we mainly
analyze one-dimension motion, but the results can be easily
extended to three-dimension situation.

1. Two kinds of description bases

As we know, contrary to classical continuous motion, the state of
quantum discontinuous motion is generally nonlocal, and the
particle undergoing this new motion moves throughout the whole
space with a certain position measure density $\rho(x,t)$ during
infinitesimal time interval, thus there exist two kinds of
description bases for quantum discontinuous motion in essence, one
should be the position of particle, which is local description
basis; the other will be the corresponding nonlocal description
basis, in the following we will demonstrate that it is just the
momentum of particle.

In fact, in order to find the nonlocal description basis, we only
need to analyze the simplest situation of the free evolution of
the new motion, where $\rho$(x,t) is constant during the
evolution, for this situation we can easily find that $j(x,t)$ is
also constant, and we have:
\begin{equation}\frac{\partial{\rho}(x,t)}{\partial{t}}=0\end{equation}
\begin{equation}\frac{\partial{j}(x,t)}{\partial{x}}=0\end{equation}
\begin{equation}\frac{\partial{j}(x,t)}{\partial{t}}=0\end{equation}
\begin{equation}\frac{\partial{\rho}(x,t)}{\partial{x}}=0\end{equation}
but the mathematical and physical meaning of these four equations
need to be analyzed, firstly, these equations can also hold in
classical fluid mechanics, but as to describing the evolution of
the new motion of a particle, their meaning will be very
different, according to these four equations, the quantities
position measure density $\rho(x,t)$ and position measure fluid
density $j(x,t)$ will be constant irrelevant to both x and t,
generally we can let $\rho(x,t)$=1, then we have
$j(x,t)$=$\rho(x,t)\cdot{v_0}$=$v_0$=$p_0$/m, where $m$ is the
mass of the particle, these two results mean that for the free
particle undergoing the new motion with one constant momentum, its
position will spread throughout the whole space with the same
position measure density, thus we demonstrate that the momentum of
particle is just the nonlocal description basis, and the momentum
state of new motion is completely nonlocal.

2. One-to-one relation

Now we have shown there are two kinds of description bases for
quantum discontinuous motion, but it is evident that there exists
only one definite motion state at any instant, so the state
description using these two kinds of description bases should be
equivalent, this means that there exists a one-to-one relation
between these two descriptions, and this relation is irrelevant to
the concrete motion state, in the following we will mainly discuss
how to find this one-to-one relation, and our analysis will also
show that this relation essentially determines the distinct
evolution for quantum discontinuous motion, as well as the axiom
framework of Hilbert space for quantum mechanics.

Like the measure density $\rho(x,t)$ and measure fluid density
$j(x,t)$ for local position x of the particle, we can also define
the measure density $f({p},t)$ and measure fluid density
$J({p},t)$ for the nonlocal momentum p of the particle, and
according to the above analysis there should exists a one-to-one
relation between the local position description $(\rho,j)$ and
nonlocal momentum description $(f,J)$. First, it is evident that
there exists no direct one-to-one relation between the measure
density functions $\rho(x,t)$ and $f(p,t)$, since even for the
above simplest situation, we have $\rho(x,t)=1$ and
$f(p,t)=\delta^2(p-p_0)$( this result can be directly obtained
when considering the general normalization relation
$\int_{\Omega}\rho(x,t)dx=\int_{\Omega}f(p,t)dp$), and there is no
one-to-one relation between them.

Then in order to obtain the one-to-one relation, we have to create
new properties on the basis of the above position description
$(\rho,j)$ and momentum description $(f,J)$, this needs a little
more mathematical trick, here we only give the main clues and the
detailed mathematical demonstrations are omitted, first, we
disregard the time variable $t$ and let $t=0$, as to the above
free evolution state with one momentum, we have
$(\rho,j)=(1,p_{0}/m)$ and $(f,J)=(\delta^2(p-p_0),0)$, thus we
need to create a new position state function $\psi(x,0)$ using $1$
and $p_{0}/m$, a new momentum state function $\varphi(p,0)$ using
$\delta^2(p-p_0)$ and $0$, and find the one-to-one relation
between these two state functions, this means there exists an
one-to-one transformation between the state functions $\psi(x,0)$
and $\varphi(p,0)$, we generally write it as follows:
\begin{equation}\label{}
\psi(x,0)=\int_{-\infty}^{+\infty}\varphi(p,0)T(p,x)dp
\end{equation}
where $T(p,x)$ is called transformation function and generally
continuous and finite for finite $p$ and $x$, since the function
$\varphi(p,0)$ will contain some form of the basic element
$\delta^2(p-p_0)$, normally we may expand it as
$\varphi(p,0)=\sum_{i=1}^{\infty}a_{i}\delta^{i}(p-p_0)$, while
the function $\psi(x,0)$ will contain the momentum $p_0$, and be
generally continuous and finite for finite $x$, then it is evident
that the function $\varphi(p,0)$ can only contain the term
$\delta(p-p_0)$, because the other terms will result in
infiniteness.

On the other hand, since the result $\varphi(p,0)=\delta(p-p_0)$
implies that there exists the relations
f(p,0)=$\varphi(p,0)^{*}\varphi(p,0)$ and
$\rho(x,0)$=$\psi(x,0)^{*}\psi(x,0)$, we may let
$\psi(x,0)=e^{iG(p_{0},x)}$ and have $T(p,x)=e^{iG(p,x)}$, then
considering the symmetry between the properties position and
momentum( this symmetry essentially stems from the equivalence
between these two kinds of descriptions, the direct implication is
for $\rho(x,0)=\delta^2(x-x_0)$ we also have $f(p,0)=1$ ) we have
the extension $G(p,x)=\sum_{i=1}^{\infty}b_{i}(px)^{i}$, but the
symmetry between the properties position and momentum further
results in the symmetry between the transformation $T(p,x)$ and
its reverse transformation $T^{-1}(p,x)$, where $T^{-1}(p,x)$
satisfies the relation
$\varphi(p,0)=\int_{-\infty}^{+\infty}\psi(x,0)T^{-1}(p,x)dp$,
thus we can only have the term $px$ in the function $G(p,x)$, for
this situation the symmetry relation between these two
transformations is $T^{-1}(p,x)=T^{*}(p,x)=e^{-ipx}$, and we let
$b_{1}=1/\hbar$, where $\hbar$ is a constant quantity, for
simplicity we let $\hbar=1$ in the following discussions. Then
mainly owing to the essential symmetry involved in the new motion
we work out the basic one-to-one relation, it is
$\psi(x,0)=\int_{-\infty}^{+\infty}\varphi(p,0)e^{ipx}dp$, where
$\psi(x,0)=e^{-ip_{0}x}$ and $\varphi(p,0)=\delta(p-p_0)$.

In fact, there may exist other complex forms for the state
functions $\psi(x,0)$ and $\varphi(p,0)$, for example, they are
not the above simple number functions but multidimensional vector
functions such as
$\psi(x,0)=(\psi_{1}(x,0),\psi_{2}(x,0),...,\psi_{n}(x,0))$ and
$\varphi(p,0)=(\varphi_{1}(p,0),\varphi_{2}(p,0),...,\varphi_{n}(p,0))$,
but the above one-to-one relation still exists for every component
function, and these vector functions still satisfy the above
modulo square relations, namely
$\rho(x,0)=\sum_{i=1}^{n}\psi_{i}(x,0)^{*}\psi_{i}(x,0)$ and
$f(p,0)=\sum_{i=1}^{n}\varphi_{i}(p,0)^{*}\varphi_{i}(p,0)$, these
complex forms will correspond to more complex theories, say,
involving more inner properties of the particle such as charge and
spin etc.

Now, since the one-to-one relation between the position state
description and momentum state description is irrelevant to the
concrete motion state of the new motion, the above one-to-one
relation for the free motion state with one momentum should apply
for any motion state of the new motion,
and the states which satisfy the one-to-one relation will be the
possible motion states of the new motion. Furthermore, it is
evident that this one-to-one relation will directly result in the
famous uncertainty relation $\triangle x\cdot\triangle
p\geq\hbar/2$, and as we will demonstrate, it will essentially
result in the consistent axiom system of quantum mechanics.

3. Axiom I of quantum mechanics

Now, the direct linear superposition of the above momentum state
$\psi_{p}(x,0)=e^{-ipx}$, which can be called momentum eigenstate,
will evidently satisfy the one-to-one relation, and should be the
possible motion state of the new motion, at the same time, Fourier
analysis further shows that any normal motion state can be
expanded as the linear superposition of the momentum eigenstates,
then all normal motion states of the new motion will form an
abstract complex linear space, which has been called Hilbert
space, thus we have shown that any motion state of the system
undergoing the new motion will correspond to the state vector in
Hilbert space, namely the Axiom I of quantum
mechanics\cite{Neumann}\cite{Jammer}, for example, the free new
motion state with two momenta corresponds to the state vector in
Hilbert space, which is a certain kind of linear superposition of
two free new motion states with one momentum in this space.

4. Axiom II of quantum mechanics

From a mathematical point of view, the above one-to-one relation
will also physically determine the linear operator structure in
Hilbert space, which may give a more abstract but deeper
description of the new motion, thus the Axiom II of quantum
mechanics\cite{Neumann}\cite{Jammer} is further included, namely,
every observable of the system corresponds to the self-adjoint
operator in the Hilbert state space, the self-adjoint requirement
guarantees the real value of the corresponding physical
observable, which may be the position observable $x$, the momentum
observable $p$ and the energy observable $E$ etc, and in general
the relations of these observables can be extended to the
corresponding operators.

Furthermore, through defining the linear operators $\hat{x}$ and
$\hat{p}$, the above one-to-one relation describing the new motion
will result in the famous noncommuting relation
$[\hat{x},\hat{p}]=i\hbar$, which is taken as the basis for
quantizing anything.

5. Axiom III of quantum mechanics

Now, we will demonstrate that the Axiom III of quantum
mechanics\cite{Neumann}\cite{Jammer}, as well as the
irreducibility of probability in quantum mechanics, results from
the objective nature of the quantum discontinuous motion, first,
the proper measurement in physics should reflect the property of
the measured system as truly as possible, and it is also rational
to presuppose the existence of such proper measurement for the new
motion, as well as for other realities; secondly, if we assume
that each measurement about the system undergoing the new motion
will bring about only one definite result, then the objective
measure density $\rho$ of the new motion will naturally result in
the probability distribution of the measurement results, which is
$P=|\psi|^2$ for discrete observable and
$P=\int_{E_1}^{E_2}|\psi|^2dE$ for continuous observable, where
$[E_1,E_2]$ is the result interval, this is just the Axiom III of
quantum mechanics. Furthermore, there does not exist one kind of
point description for the quantum discontinuous motion in essence,
so the probability for a single definite measurement result is
irreducible, it is essentially determined by the discontinuity and
value-dispersion nature of the quantum discontinuous motion.

6. Axiom IV of quantum mechanics

Now, we will work out the dynamical evolution law of the new
motion, namely the Axiom IV of quantum mechanics.

First, in order to find how the time variable $t$ is included in
the functions $\psi(x,t)$ and $\varphi(p,t)$, we may consider the
linear superposition of two momentum eigenstates, namely
$\psi(x,t)=\frac{1}{\sqrt{2}}[e^{ip_{1}x-ic_{1}(t)}+e^{ip_{2}x-ic_{2}(t)}]$,
then the position measure density is $\rho(x,t)=[1+\cos(\triangle
c(t)-\triangle px)]/2$, where $\triangle c(t)=c_{2}(t)-c_{1}(t)$
and $\triangle p=p_{2}-p_{1}$, now we let $\triangle p\rightarrow
0$, then we have $\rho(x,t)\rightarrow 1$ and $\triangle
c(t)\rightarrow 0$, especially using the conservation relation we
can get $dc(t)/dt=dp\cdot p/m$, namely $dc(t)=d(p^{2}/m)\cdot t$
or $dc(t)=dE\cdot t$, where $E=p^{2}/m$, is the energy of the
particle in the nonrelativistic domain, thus as to any momentum
eigenstate we have the time-included formula
$\psi(x,t)=e^{ipx-iEt}$.


Now, as to the free motion state with one momentum, namely the
momentum eigenstate $\psi(x,t)=e^{ipx-iEt}$, using the
nonrelativistic relation $E=\frac{p^{2}}{m}$ and including the
constant quantity $\hbar$ we can easily find its nonrelativistic
evolution law, which is
\begin{equation}
i\hbar\frac{\partial{\psi}(x,t)}{\partial{t}}=
-\frac{\hbar^2}{2m}\cdot{\frac{\partial^{2}\psi(x,t)}{\partial{x^2}}}
\end{equation}
then owing to the linearity of this equation, this evolution
equation also applies to the linear superposition of the momentum
eigenstates, namely all possible free notion states, or we can
say, it is the free evolution law for the new motion; Secondly, we
will consider the evolution law for the new motion under outside
potential, when the potential $U(x,t)$ is a constant $U$, the
evolution equation will be
\begin{equation}
i\hbar\frac{\partial{\psi}(x,t)}{\partial{t}}=
-\frac{\hbar^2}{2m}\cdot{\frac{\partial^{2}\psi(x,t)}{\partial{x^2}}}
+U\cdot{\psi(x,t)}
\end{equation}
then when the potential $U(x,t)$ is related to x and t, the above
equation will also essentially determine the evolution equation,
it is
\begin{equation}
i\hbar\frac{\partial{\psi}(x,t)}{\partial{t}}=
-\frac{\hbar^2}{2m}\cdot{\frac{\partial^{2}\psi(x,t)}{\partial{x^2}}}
+U(x,t)\cdot{\psi(x,t)}
\end{equation}
for three-dimension situation the equation will be
\begin{equation}
i\hbar\frac{\partial{\psi}(\mathbf{x},t)}{\partial{t}}=
-\frac{\hbar^2}{2m}\cdot\nabla^{2}\psi(\mathbf{x},t)
+U(\mathbf{x},t)\cdot{\psi(\mathbf{x},t)}
\end{equation}
this is just the Schr\"{o}dinger equation in quantum mechanics,
thus we have deduced the Axiom IV of quantum
mechanics\cite{Neumann}\cite{Jammer}.

On the other hand, according to the Axiom II of quantum mechanics,
every observable of the system corresponds to a self-adjoint
operator in the Hilbert state space, and in general the relations
of these observables can be extended to the corresponding
operators, thus the above evolution equation of new motion can be
reduced to the familiar nonrelativistic energy equality
$E=p^2/m+U$, and the equation can be rewritten as
$\hat{E}\psi(\mathbf{x},t)=\hat{p}^{2}/m\psi(\mathbf{x},t)+U(\hat{x},t)\psi(\mathbf{x},t)$,
where $\hat{E}=i\hbar\partial/\partial t$, and
$\hat{p}=-i\hbar\nabla$.

At last, the above analysis also shows that the state function
$\psi(x,t)$ provides a complete description of the quantum
discontinuous motion, since the new motion is completely described
by the measure density $\rho(x,t)$ and measure fluid density
$j(x,t)$, and according to the above evolution equation the state
function $\psi(x,t)$ can be formulated by these two functions,
namely $\psi(x,t)=\rho^{1/2}\cdot{e^{iS(x,t)/\hbar}}$, where
$S(x,t)=m\int_{-\infty}^{x} j(x^{'},t)/\rho(x^{'},t)dx^{'}$( Note:
when in three-dimension space, the formula for S(x,y,z,t) will be
$S(x,y,z,t)=m\int_{-\infty}^{x}
j_{x}(x^{'},y,z,t)/\rho(x^{'},y,z,t)dx^{'}=m\int_{-\infty}^{y}
j_{y}(x,y^{'},z,t)/\rho(x,y^{'},z,t)dy^{'}=m\int_{-\infty}^{z}
j_{z}(x,y,z^{'},t)/\rho(x,y,z^{'},t)dz^{'}$, since in general
there exists the relation $\nabla \times
\{\mathbf{j}(x,y,z,t)/\rho(x,y,z,t)\}=0$ when $\rho(x,y,z,t) \neq
0$), and these two functions can also be expressed by the state
function, namely $\rho(x,t)=|\psi(x,t)|^2$ and $j(x,t) = [\psi^*
\partial \psi/\partial t - \psi \partial
 \psi^*/\partial t ]/2i$, thus there exists a one-to-one relation
between $(\rho(x,t),j(x,t))$ and $\psi(x,t)$, and the state
function $\psi(x,t)$ also provides a complete description of the
quantum discontinuous motion.

On the other hand, we can see that the absolute phase of the wave
function $\psi(x,t)$, which may depends on time in nonrelativistic
domain, is useless for describing the new motion, since according
to the above analysis it disappears in the measure density
$\rho(x,t)$ and measure fluid density $j(x,t)$, which completely
describe the quantum discontinuous motion, thus from the point of
view of the new motion it is natural that the absolute phase of
the wave function possesses no physical meaning.

7. Axiom V of quantum mechanics

As we know, quantum mechanics is self-consistent when defined by
the above four axioms, since the elucidation about the
corresponding relation between the physical reality and
mathematical language is just a conditional statement, namely if
the measurement about the quantum system brings about only one
definite result, then the probability distribution of the
measurement results will satisfy the formula, say for the discrete
observable, $P=|\psi|^2$; but if only the above four axioms are
included, quantum mechanics will be evidently incomplete, since it
does not describe the real measurement result, thus its founders
resorted to the projection postulate or Axiom
V\cite{Neumann}\cite{Jammer} in order to account for the
measurement process, whereas this postulate is still a direct
description about the measurement result, it says nothing about
how the measurement can and does bring about one definite result,
so it needs to be further explained in physics, now the new motion
of particle just provide such a broad framework for objectively
studying the microscopic world that it may solve the collapse
problem, for example, there may exist many kinds of concrete
motion modes among the new motion, and the new motion may display
differently in the nonrelativistic and relativistic domains,
especially when involving gravity the new motion of particle may
naturally result in the objective collapse process, but owing to
the weird difficulty of this problem, we will tackle it in another
paper.

8. The value of $\hbar$ in quantum mechanics

Up to now, one problem is still left, it is how to determine the
value of $\hbar$ in quantum mechanics or our world, according to
the above analysis we only know that the constant $\hbar$
possesses a finite nonzero value, certainly, just like the other
physical constants such as c and G, its value can be determined by
the experience, but its existence need to be explained, and the
above analysis about the new motion can provide the answer, namely
the existence of $\hbar$ essentially results from the
irreducibility of the nonlocal momentum definition, or
nonexistence of the velocity or local momentum of the particle
undergoing the new motion, this kind of irreducibility denotes
that momentum is no longer related to space-time or velocity as
for classical continuous motion, it is also an essential property
of the new motion just like position, especially it provides an
equivalent nonlocal description of the new motion, while position
provides a local description of the new motion, this equivalence
further results in the one-to-one relation between these two kinds
of descriptions, then it is just this one-to-one relation which
requires the existence of a certain constant $\hbar$ to cancel out
the unit of the physical quantities $px$ and $Et$ in the relation,
at the same time, the existence of $\hbar$ also indicates some
kind of balance between the properties( concretely speaking, their
value distribution dispersions ) limited by the one-to-one
relation ( there is no such limitation for classical continuous
motion and $\hbar=0$ ), or we can say, the existence of $\hbar$
essentially indicates some kind of balance between the nonlocality
and locality of the new motion in space-time.

Certainly, the new motion provides a broader motion framework for
the particle in microscopic world, in which we can understand the
weird displays of the microscopic objects objectively and
consistently, which can not be grasped consistently in the old
framework of classical continuous motion, but it can not give the
concrete value of $\hbar$ in our world by itself, as special
relativity can not determine the value of light velocity $c$, or
general relativity can not determine the value of gravity constant
$G$, surely, there may exist some deeper reasons for the
particular value of $\hbar$ in our universe, but the new motion
can not determine this value alone, the solution may have to
resort to other subtle realities in this world, for example,
gravity (G), space-time(c), or even the existence of mankind.

\section{The confirmation of the new motion}
In the following, we will give two main methods to confirm the new
motion underlying the wave function in microscopic world.

\subsection{Protective measurement}
The first method is called protective
measurement\cite{Aha1}\cite{Aha2}, it aims at measure the new
motion state of a single particle through repeatedly measuring it
without destroying its state, in real experiment a small ensemble
of similar particles may be required. By use of this kind of
measurement, the new motion state or wave function of a particle
does not change appreciably when the measurement is being made on
it, its clever way is to let the system undergo a suitable
interaction so that it is in a non-degenerate eigenstate of the
whole Hamiltonian, then the measurement is made adiabatically so
that the new motion state described by the wave function neither
changes appreciably nor becomes entangled with the measurement
device, this suitable interaction is called the protection.

In the following, we will demonstrate how to use the protective
measurement to confirm the new motion of a single particle in
microscopic world, which is described by the wave function and
Schr\"{o}dinger equation, for simplicity but lose no generality,
we only consider a particle in a discrete nondegenerate energy
eigenstate $\psi(x)$, the interaction Hamiltonian for measuring
the value of an observable $A_n$ in this state is:$H = g(t) P A_n
$, which couples the system to a measuring device, with coordinate
and momentum denoted respectively by $Q$ and $P$, where $A_n$ is
the normalized projection operator on small regions $V_n$ having
volume $v_n$, namely:

$$A_n=\cases{ {1\over {v_n}},&if $x \in V_n$,\cr 0,&if $x \not\in
V_n$.\cr} \eqno(3) $$ \noindent

the time-dependent coupling $g(t)$ is normalized to $\int_{0}^{T}
g(t) dt =1$, we let $g(t) = 1/T$ for most of the time $T$ and
assume that $g(t)$ goes to zero gradually before and after the
period $T$ to obtain an adiabatic process when $T \rightarrow
\infty$, the initial state of the pointer is taken to be a
Gaussian centered around zero, and the canonical conjugate $P$ is
bounded and also a motion constant not only of the interaction
Hamiltonian, but of the whole Hamiltonian.

Now using this kind of protective measurement, the measurement of
$A_n$ yields the result:

$$\langle A_n \rangle = {1\over {v_n}} \int_{V_n} |\psi|^2 dv =
|\psi_n|^2 $$

the result $ |\psi_n|^2 $ is just the average of the measure
density $\rho(\mathbf{x}) = |\psi(\mathbf{x})|^2$ over the small
region $V_n$, so when $v_n \rightarrow 0$ and after performing
measurements in sufficiently many regions $V_n$ we can find the
measure density $\rho(\mathbf{x})$ of the new motion state of the
measured particle.

Then we will measure the measure current density
$\mathbf{j}(\mathbf{x})$ of the new motion state, namely we need
measure the value of an observable $B_n$ in this state, where $B_n
={1\over{2i}} (A_n\nabla + \nabla A_n)$, the measurement result
will be $\langle B_n\rangle$, ant it is just the average value of
the measure fluid density $\mathbf{j}(\mathbf{x}) = {1\over{2i}}
(\psi^* \nabla \psi - \psi \nabla \psi^* )$ in the region $V_n$,
so when $v_n \rightarrow 0$ and after performing measurements in
sufficiently many regions $V_n$, we can also find the measure
fluid density $\mathbf{j}(\mathbf{x})$ of the new motion state of
the measured particle.

Thus we have demonstrated that the new motion of a single
particle, which is described by the measure density
$\rho(\mathbf{x})$ and measure fluid density
$\mathbf{j}(\mathbf{x})$, or the abstract wave function
$\psi(\mathbf{x})$, can be confirmed through the above protective
measurement.

\subsection{Standard impulse measurement}
Certainly, the standard impulse measurement in quantum
mechanics\cite{Neumann} can also confirm the new motion of a
single particle described by the wave function in microscopic
world, its wrinkle lies in that we first prepare an ensemble of a
large number of particles in the same state of new motion, then
using this kind of measurement measure every particle in the
ensemble only one time, and we need not repeatedly measure the
same particle, thus even if the state of the new motion or wave
function of a single particle will be destroyed after each
measurement so that the following measurement will no longer
reveal the real information about the original state of new
motion, but according to quantum mechanics, all the individual
measurement results about the ensemble can reveal the state of the
ensemble, and also the state of the new motion of a single
particle in the ensemble, since every particle in the ensemble is
in the same state of new motion.

\subsection{Understanding the wave function in terms of new motion}
As we know, many ontological interpretations of quantum mechanics
still consider the wave function as one kind of objective field,
as Bell said, "No one can understand this theory until he is
willing to think of $\psi$ as a real objective field rather than a
'probability amplitude'"\cite{Bell}, but the difficulties involved
in this kind of ontological interpretation had been pointed out
since the beginning of quantum mechanics, for example, the
existence of complex wave function, the multidimensionality of the
wave function and the representation problem etc can not be solved
in the framework of objective field, these difficulties greatly
prevented physicists from accepting the objective view about the
microscopic object described by the wave function. Now, according
to the new quantum discontinuous motion, we can easily overcome
these difficulties in the framework of objective particle:

(1). As to the problem of complex wave function, since the wave
function is not one kind of objective field at all, it is just one
kind of indirect abstract mathematical symbol, which is used to
describe the objective quantum discontinuous motion of the
particles in microscopic world, concretely speaking, it is an
abstract complex of the measure density $\rho({x},t)$ and measure
fluid density ${j}({x},t)$, which directly describe the state of
the new motion in physics, and its appearance essentially results
from the symmetry involved in the new motion and the resulting
linear evolution principle, thus whether the wave function is
complex or not is not a problem for the new motion described by
the wave function.

(2). The multidimensionality of the wave function is very natural
from the point of view of new motion of particle, as we know, the
measure density $\rho(\mathbf{x},t)$, or wave function
$\psi({\mathbf{x}},t)$ for a single particle depends on three
space variables, and as to two particles generally we can not
define their respective measure densities
$\rho_{1}({\mathbf{x}},t)$ and $\rho_{2}({\mathbf{x}},t)$, or wave
functions $\psi_{1}({\mathbf{x}},t)$ and
$\psi_{2}({\mathbf{x}},t)$, whereas we should define their joint
measure density $\rho({\mathbf{x}}_{1},{\mathbf{x}}_{2},t)$
according to point set theory, which means the joint measure
density of particle 1 in position ${\mathbf{x}}_1$ and particle 2
in position ${\mathbf{x}}_2$, this further results in that the
one-to-one relation is two-manifold Fourier transformation
involving the wave function
$\psi({\mathbf{x}}_{1},{\mathbf{x}}_{2},t)$, which depends on six
space variables, namely
\begin{equation}
\psi({\mathbf{x}}_{1},{\mathbf{x}}_{2},t)=\int_{-\infty}^{+\infty}\int_{-\infty}^{+\infty}
\varphi({\mathbf{p}}_{1},{\mathbf{p}}_{2},t)
e^{i({\mathbf{x}}_{1}\cdot{\mathbf{p}}_{1}+{\mathbf{x}}_{2}\cdot{\mathbf{p}}_{2})}d{\mathbf{p}}_{1}d{\mathbf{p}}_{2}
\end{equation}
and Schr\"{o}dinger equation for this two-particle situation is
\begin{equation}
i\hbar\frac{\partial{\psi}({\mathbf{x}}_{1},{\mathbf{x}}_{2},t)}{\partial{t}}
=- \frac{\hbar^2}{2m}[\nabla^{2}_{1}
+\nabla^{2}_{2}]\psi({\mathbf{x}}_{1},{\mathbf{x}}_{2},t)+U({\mathbf{x}}_{1},{\mathbf{x}}_{2},t)\cdot{\psi({\mathbf{x}}_{1},{\mathbf{x}}_{2},t)}
\end{equation}
Certainly, when these two particles are independent, the joint
measure density $\rho({\mathbf{x}}_{1},\mathbf{x}_{2},t)$ can be
reduced to
$\rho_{1}({\mathbf{x}}_{1},t)\rho_{2}(\mathbf{x}_{2},t)$, and the
joint wave function $\psi({\mathbf{x}}_{1},{\mathbf{x}}_{2},t)$
can also be reduced to
$\psi_{1}({\mathbf{x}}_{1},t)\psi_{2}(\mathbf{x}_{2},t)$.

Certainly, as Bohr had taught us, if one does not wander about
quantum mechanics he surely has not understood this theory yet.
Indeed, the quantum discontinuous motion in microscopic world is
extremely different from the familiar classical continuous motion
in macroscopic world, so it is natural that the experience from
classical mechanics contradicts the picture of new motion,
especially the particle undergoing the new motion can move far
away in a very small time interval, or even infinitesimal time
interval, which appears to conflict with the local spreading of
energy, but in fact, this just provides the objective origin of
quantum nonlocality, which has been confirmed in experiments; on
the other hand, when confirming this kind of weird display of new
motion we have to consider the more weird quantum measurement,
which needs to be further studied in another paper, but here we
should keep in mind something, namely when we go into the new
field of quantum discontinuous motion, we should re-understand the
meaning of everything in principle, including energy and the
accepted law that it moves locally or even the concept of
particle, if we still use them, in one word, our understanding
about reality can only be determined by the reality itself, not
our belief.

\section{Explaining the weird display of the wave function in terms of new motion}
At last, we will give two familiar examples, which have evidently
manifested the existence of new motion in microscopic world, to
explain the weird displays of the wave function in terms of the
new motion of particle.

The first is the base state of Hydrogen atom, its position
distribution density, which can be found through the above
measurements, is written in the
following:\begin{equation}\rho(\mathbf{x})=|\psi(\mathbf{x})|^2=\frac{4}{a_0}\cdot{exp(-
\frac{2r}{a_0})}\end{equation}

According to the new motion of the particle, at any instant the
electron will be in only one position in space, but during
infinitesimal time interval [t,t+dt], the electron will move
throughout the whole space where the above function does not equal
to zero, and its position measure density will be the same as the
above position distribution density function obtained from the
wave function, thus during infinitesimal time interval [t,t+dt],
according to Gauss theorem the charge distribution of the whole
system will be equivalent to the zero charge distribution for the
outside observer, and there exists no change of the whole charge
distribution either, so it can be easily understood that no energy
is radiated during finite time interval, as well as during
infinitesimal time interval, this is just the objective origin of
the mysterious stability in atom world.

The second example is the double-slit experiment, people have been
trying to understand the formation process of the double-slit
interference pattern objectively, but few people can give an
ontological description for it up to now, the essential reason, as
we think, is that people all ignored the difference between
instant and infinitesimal time interval. By means of the new
quantum discontinuous motion of the particle, the mystery of this
process can be disclosed, the real process should be that the
particle undergoing the new motion passes through both slits in
the double-slit experiment, this means that the particle is still
in only one of the two slits at any instant, but during the time
interval $\Delta t $, which can approach to zero, the particle
moves throughout both slits and passes through them, and the
position measure density of the particle always satisfies the
function $\rho(\mathbf{x},t)=|\psi(\mathbf{x},t)|^2$, which is
finite and the same in both slits. Since the particle undergoing
the new motion can pass through both slits in this objective way,
we can more easily understand the forming of double-slit
interference pattern, which is not a simple superposition of two
one-slit interference patterns.

\section{Conclusions}
On the whole, a new point of view about the motion of the particle
is presented to interpret the weird display of the wave function
describing the microscopic object, in mathematics, the point set
theory casts a new light on the study of physical motion,
especially through deeply analyzing the regular dense point set,
we present a new kind of motion, which is called quantum
discontinuous motion contrary to classical continuous motion. Then
the physical meaning of this new motion is carefully examined, at
the same time, we give a strict physical definition about the
quantum discontinuous motion, and show that the notorious wave
function is just a mathematical complex describing the new motion
of the microscopic particle, and Schr\"{o}dinger equation in
quantum mechanics is just the simplest nonrelativistic evolution
equations for the new motion, especially the consistent axiom
system of quantum mechanics is also deduced out. At last, the
protective measurement and standard impulse measurement are used
to confirm the existence of the new motion of the microscopic
particle described by the wave function, and two famous examples
are also given to explain the weird display of the wave function
in terms of new motion.

\vskip .5cm \noindent Acknowledgments \vskip .5cm Thanks for
helpful discussions with X.Y.Huang ( Peking University ),
A.Jadczyk ( University of Wroclaw ), P.Pearle ( Hamilton College
), F.Selleri ( University di Bari ), Y.Shi ( University of
Cambridge ), A.Shimony, A.Suarez ( Center for Quantum Philosophy
), L.A.Wu ( Institute Of Physics, Academia Sinica ), Dr S.X.Yu (
Institute Of Theoretical Physics, Academia Sinica ), H.D.Zeh.

\end{document}